\documentclass[twocolumn]{revtex4}
\usepackage{graphicx}
\begin{document}
\title{Constraints on Neutrino Oscillations Using
1258 Days of Super-Kamiokande Solar Neutrino Data}

\newcounter{foots}
\newcounter{notes}
\newcommand{\authoraticrr}{$^{1}$}
\newcommand{\authoratbu}{$^{2}$}
\newcommand{\authoratbnl}{$^{3}$}
\newcommand{\authoratuci}{$^{4}$}
\newcommand{\authoratcsu}{$^{5}$}
\newcommand{\authoratgmu}{$^{6}$}
\newcommand{\authoratgifu}{$^{7}$}
\newcommand{\authoratuh}{$^{8}$}
\newcommand{\authoratkek}{$^{9}$}
\newcommand{\authoratkobe}{$^{10}$}
\newcommand{\authoratkyoto}{$^{11}$}
\newcommand{\authoratlanl}{$^{12}$}
\newcommand{\authoratlsu}{$^{13}$}
\newcommand{\authoratlsuumd}{$^{13,14}$}
\newcommand{\authoratumd}{$^{14}$}
\newcommand{\authoratduluth}{$^{15}$}
\newcommand{\authoratsuny}{$^{16}$}
\newcommand{\authoratniigata}{$^{17}$}
\newcommand{\authoratosaka}{$^{18}$}
\newcommand{\authoratseoul}{$^{19}$}
\newcommand{\authoratshizuokasc}{$^{20}$}
\newcommand{\authoratshizuoka}{$^{21}$}
\newcommand{\authorattohoku}{$^{22}$}
\newcommand{\authorattokyo}{$^{23}$}
\newcommand{\authorattokai}{$^{24}$}
\newcommand{\authorattit}{$^{25}$}
\newcommand{\authoratwarsaw}{$^{26}$}
\newcommand{\authoratuw}{$^{27}$}

\newcommand{\addressoficrr}[1]{$^{1}$ #1 }
\newcommand{\addressofbu}[1]{$^{2}$ #1 }
\newcommand{\addressofbnl}[1]{$^{3}$ #1 }
\newcommand{\addressofuci}[1]{$^{4}$ #1 }
\newcommand{\addressofcsu}[1]{$^{5}$ #1 }
\newcommand{\addressofgmu}[1]{$^{6}$ #1 }
\newcommand{\addressofgifu}[1]{$^{7}$ #1 }
\newcommand{\addressofuh}[1]{$^{8}$ #1 }
\newcommand{\addressofkek}[1]{$^{9}$ #1 }
\newcommand{\addressofkobe}[1]{$^{10}$ #1 }
\newcommand{\addressofkyoto}[1]{$^{11}$ #1 }
\newcommand{\addressoflanl}[1]{$^{12}$ #1 }
\newcommand{\addressoflsu}[1]{$^{13}$ #1 }
\newcommand{\addressofumd}[1]{$^{14}$ #1 }
\newcommand{\addressofduluth}[1]{$^{15}$ #1 }
\newcommand{\addressofsuny}[1]{$^{16}$ #1 }
\newcommand{\addressofniigata}[1]{$^{17}$ #1 }
\newcommand{\addressofosaka}[1]{$^{18}$ #1 }
\newcommand{\addressofseoul}[1]{$^{19}$ #1 }
\newcommand{\addressofshizuokasc}[1]{$^{20}$ #1 }
\newcommand{\addressofshizuoka}[1]{$^{21}$ #1 }
\newcommand{\addressoftohoku}[1]{$^{22}$ #1 }
\newcommand{\addressoftokyo}[1]{$^{23}$ #1 }
\newcommand{\addressoftokai}[1]{$^{24}$ #1 }
\newcommand{\addressoftit}[1]{$^{25}$ #1 }
\newcommand{\addressofwarsaw}[1]{$^{26}$ #1 }
\newcommand{\addressofuw}[1]{$^{27}$ #1 }

\author{
{\large The Super-Kamiokande Collaboration} \\ 
\bigskip
S.~Fukuda\authoraticrr,
Y.~Fukuda\authoraticrr,
M.~Ishitsuka\authoraticrr, 
Y.~Itow\authoraticrr,
T.~Kajita\authoraticrr, 
J.~Kameda\authoraticrr, 
K.~Kaneyuki\authoraticrr,
K.~Kobayashi\authoraticrr, 
Y.~Koshio\authoraticrr, 
M.~Miura\authoraticrr
% ...
\\
\smallskip
\footnotesize
\it
\addressoficrr{Institute for Cosmic Ray Research, University of Tokyo, Kashiwa,Chiba 277-8582, Japan}\\
\addressofbu{Department of Physics, Boston University, Boston, MA 02215, USA}\\
\addressofbnl{Physics Department, Brookhaven National Laboratory, Upton, NY 11973, USA}\\
\addressofuci{Department of Physics and Astronomy, University of California, Irvine, Irvine, CA 92697-4575, USA }\\
\addressofcsu{Department of Physics, California State University, Dominguez Hills, Carson, CA 90747, USA}\\
\addressofgmu{Department of Physics, George Mason University, Fairfax, VA 22030, USA }\\
\addressofgifu{Department of Physics, Gifu University, Gifu, Gifu 501-1193, Japan}\\
\addressofuh{Department of Physics and Astronomy, University of Hawaii, Honolulu, HI 96822, USA}\\
\addressofkek{Institute of Particle and Nuclear Studies, High Energy Accelerator Research Organization (KEK), Tsukuba, Ibaraki 305-0801, Japan }\\
\addressofkobe{Department of Physics, Kobe University, Kobe, Hyogo 657-8501, Japan}\\
\addressofkyoto{Department of Physics, Kyoto University, Kyoto 606-8502, Japan}\\
\addressoflanl{Physics Division, P-23, Los Alamos National Laboratory, Los Alamos, NM 87544, USA }\\
\addressoflsu{Department of Physics and Astronomy, Louisiana State University, Baton Rouge, LA 70803, USA }\\
\addressofumd{Department of Physics, University of Maryland, College Park, MD 20742, USA }\\
\addressofduluth{Department of Physics, University of Minnesota
Duluth, MN 55812-2496, USA}\\
\addressofsuny{Department of Physics and Astronomy, State University of New York, Stony Brook, NY 11794-3800, USA}\\
\addressofniigata{Department of Physics, Niigata University, Niigata, Niigata 950-2181, Japan }\\
\addressofosaka{Department of Physics, Osaka University, Toyonaka, Osaka 560-0043, Japan}\\
\addressofseoul{Department of Physics, Seoul National University, Seoul 151-742, Korea}\\
\addressofshizuokasc{International and Cultural Studies, Shizuoka Seika College, Yaizu, Shizuoka, 425-8611, Japan}\\
\addressofshizuoka{Department of Systems Engineering, Shizuoka University, Hamamatsu, Shizuoka 432-8561, Japan}\\
\addressoftohoku{Research Center for Neutrino Science, Tohoku University, Sendai, Miyagi 980-8578, Japan}\\
\addressoftokyo{The University of Tokyo, Tokyo 113-0033, Japan }\\
\addressoftokai{Department of Physics, Tokai University, Hiratsuka, Kanagawa 259-1292, Japan}\\
\addressoftit{Department of Physics, Tokyo Institute for Technology, Meguro, Tokyo 152-8551, Japan }\\
\addressofwarsaw{Institute of Experimental Physics, Warsaw University, 00-681 Warsaw, Poland }\\
\addressofuw{Department of Physics, University of Washington, Seattle, WA 98195-1560, USA}\\
}
\affiliation{ } 


\begin{abstract}
We report the result of a search for neutrino oscillations
using precise measurements of the recoil electron energy spectrum and
zenith angle variations of the solar neutrino flux
from 1258 days of neutrino-electron scattering data
in Super-Kamiokande.
The absence of significant zenith angle variation and spectrum
distortion places strong constraints on neutrino mixing and mass difference
in a flux-independent way. Using the Super-Kamiokande
flux measurement in addition, two allowed regions at large mixing
are found.
\end{abstract}

\maketitle


\begin{table*}[bt]
\caption{Flux, uncertainty and definition of
zenith angle and energy bins. The systematic uncertainty
in the last two columns is split into energy-uncorrelated
and energy-correlated uncertainty. The systematic uncertainty
is assumed to be fully correlated in zenith angle.
\label{tab:error}}
\newcommand{\plumi}[2]{\matrix{+#1\\[-1.2mm]-#2}}
\begin{center}
\begin{tabular}{l|ccccccc|cl}
\hline
& 
\multicolumn{7}{c|}{Flux$\pm$statistical uncertainty in units of SSM} &
\multicolumn{2}{c}{syst. uncert. in \%} \cr 
\hline
& 
Day             &  Mantle 1       &  Mantle 2       &  Mantle 3       &
Mantle 4        &  Mantle 5       &  Core           & 
\multicolumn{2}{c}{Energy-} \cr
cos$\theta_z$-Range    & 
-1.00--0.00     &  0.00--0.16     &  0.16--0.33     & 0.33--0.50      &
 0.50--0.67     &  0.67--0.84     &  0.84--1.00  &
uncorr.& correlated \cr
\hline
 5.0--5.5 MeV & \multicolumn{7}{c|}{0.436$\pm$0.046} & 
$\plumi{3.9}{3.1}$ & $\plumi{0.25}{0.21}$ \cr
 5.5--6.5 MeV & 
0.431$\pm$0.022 & 0.464$\pm$0.060 & 0.410$\pm$0.055 & 0.442$\pm$0.048 &
0.453$\pm$0.048 & 0.495$\pm$0.054 & 0.434$\pm$0.058 & 
$\plumi{1.5}{1.4}$ & $\plumi{0.30}{0.26}$ \cr
 6.5--8.0 MeV & 
0.461$\pm$0.013 & 0.524$\pm$0.036 & 0.506$\pm$0.033 & 0.438$\pm$0.028 &
0.466$\pm$0.027 & 0.424$\pm$0.030 & 0.409$\pm$0.033 & 
$\pm$1.4 & $\plumi{0.77}{0.75}$ \cr
 8.0--9.5 MeV &
0.437$\pm$0.014 & 0.449$\pm$0.038 & 0.482$\pm$0.036 & 0.460$\pm$0.031 &
0.503$\pm$0.031 & 0.461$\pm$0.034 & 0.439$\pm$0.037 &
$\pm$1.4 & $\pm$1.6 \cr
 9.5--11.5 MeV&
0.434$\pm$0.015 & 0.432$\pm$0.042 & 0.493$\pm$0.040 & 0.446$\pm$0.034 &
0.448$\pm$0.034 & 0.435$\pm$0.037 & 0.484$\pm$0.044 &
$\pm$1.4 & $\plumi{3.1}{2.9}$ \cr
11.5--13.5 MeV& 
0.456$\pm$0.026 & 0.496$\pm$0.071 & 0.290$\pm$0.055 & 0.394$\pm$0.053 &
0.477$\pm$0.056 & 0.439$\pm$0.061 & 0.465$\pm$0.068 &
$\pm1.4$ & $\plumi{5.5}{5.0}$ \cr
13.5--16.0 MeV&
0.482$\pm$0.056 & 0.532$\pm$0.155 & 0.775$\pm$0.171 & 0.685$\pm$0.141 & 
0.607$\pm$0.130 & 0.471$\pm$0.128 & 0.539$\pm$0.153 &
$\pm$1.4 & $\plumi{9.2}{8.3}$ \cr
16.0--20.0 MeV& \multicolumn{7}{c|}{0.476$\pm$0.149} & 
$\pm$1.4 & $\plumi{16}{14}$ \cr
\hline
\end{tabular}
\end{center}
\end{table*}

For over 30 years, measurements of the solar neutrino flux
\cite{Cl,Kam3,SAGE,Gallex,SKflux2} have been significantly below
the prediction of the Standard
Solar Models (SSMs)\cite{BP2000,SSMs}. 
Neutrino flavor oscillations, similar to those seen in
atmospheric neutrinos\cite{SKatmos}, are a
natural explanation for this discrepancy.
This type of flavor conversion is inherently
energy-dependent. 
Since Super-Kamiokande (SK) measures the
energy of the recoil electron from elastic electron-neutrino scattering,
it has sensitivity to this energy dependence.
In addition to a conversion in vacuum,
a matter-induced resonance in
the sun\cite{MSW} may sufficiently enhance the disappearance
probability of solar neutrinos even for
small neutrino mixing.
For some oscillation parameters,
matter-enhanced oscillations within the earth can lead to a
different flux during day-time than during the night.
Therefore, either a distorted energy-dependence or a zenith angle variation
of the solar neutrino flux would be considered
evidence for oscillation. In addition, the shape
of the distortion and zenith angle variation
would determine the oscillation parameters independently from uncertainties
in the SSM's flux prediction.
As a real-time solar neutrino experiment, SK can
study the solar zenith angle flux variation.

Super-Kamiokande started taking data in April, 1996.
It has since confirmed the deficit of solar neutrinos\cite{SKflux2},
measured the recoil energy spectrum\cite{SKspec}
and carried out an initial search for zenith angle variation\cite{SKdn}.
In this report, we analyze the first 1258 days of data
(May $31^{\mbox{st}}$, 1996 through October $6^{\mbox{th}}$, 2000)
using both spectral distortion and zenith angle variation.
The total number of solar 
neutrino events
above a threshold of 5 MeV of recoil electron energy
is $18464^{+677}_{-590}$.
The resulting flux of $^8$B solar neutrinos,
$(2.32^{+0.09}_{-0.08})\times10^6/$(cm$^2$s)\cite{SKflux2},
is $0.451^{+0.017}_{-0.015}$ of the flux predicted by the reference
SSM(BP2000)\cite{BP2000}.

\begin{figure}[tb]
\centerline{\includegraphics[width=9cm]{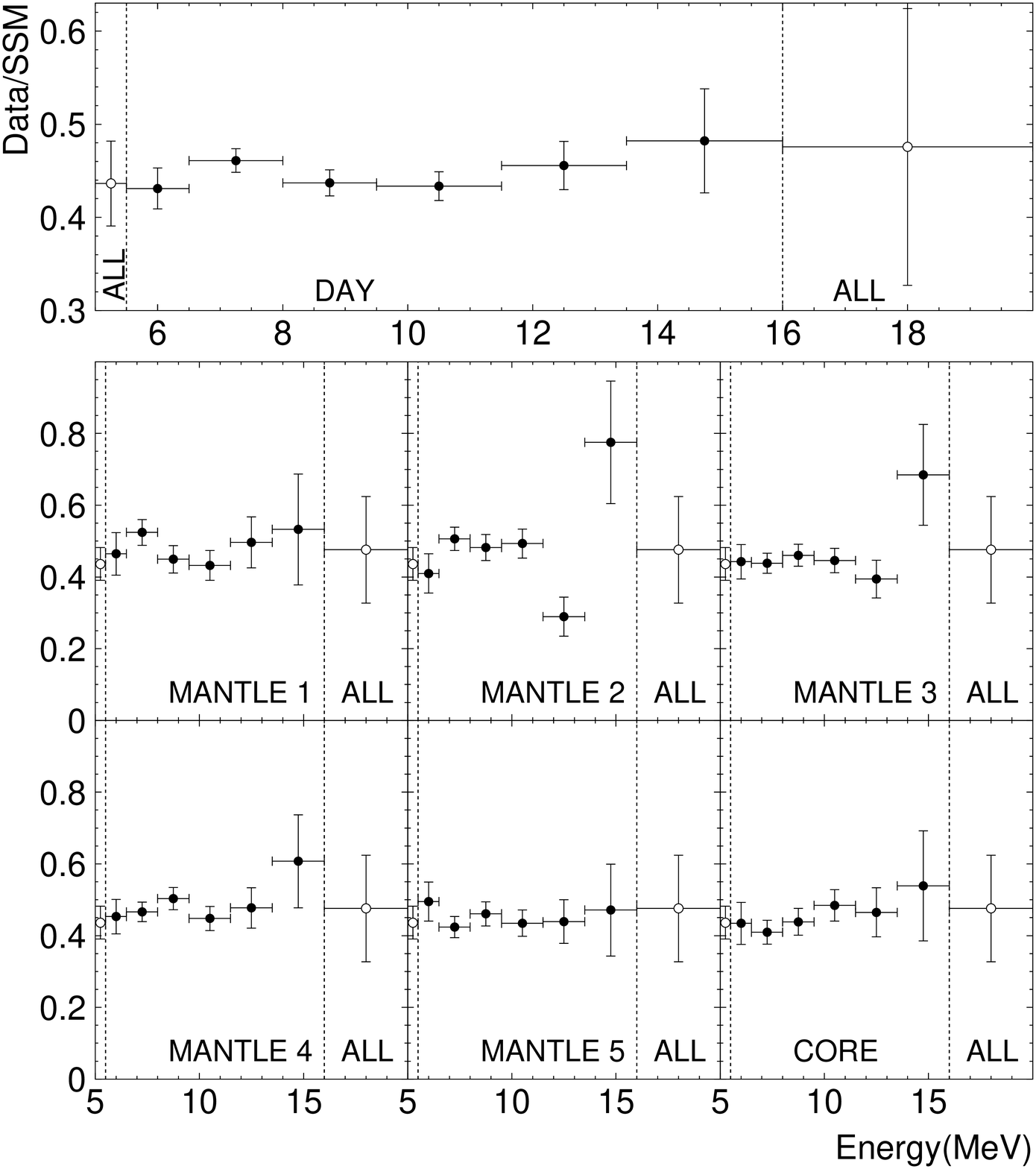}}
\caption{Spectrum between 5 and 20 MeV for various zenith angles. Data points
with open circles combine all zenith angle bins. 
The error bar is the statistical error. See table~\protect\ref{tab:error} for
systematic errors and the definition of the binnig.}
\label{daynightspec}
\end{figure}

The sample is divided into seven zenith angle bins (one day bin and
six bins in $\cos\theta_z$ for the night); within each zenith angle
bin, the data are divided into eight recoil electron bins.  We will
refer to this binning of the data as the ``zenith angle spectrum''
(see Fig.~\ref{daynightspec}).  We define the zenith angle $\theta_z$
of an event as the angle between the vertical direction and the solar
direction at the time of the event. Day events have $\cos
\theta_z\le0$ and night events $\cos \theta_z>0$.  The size of the
sample (already divided into seven zenith angles) does not allow a
subdivision into 19 energy bins shown in~\cite{SKflux2}. Due to this
statistical limitation the lowest (5.0-5.5 MeV) and the highest
(16.0-20.0 MeV) energy bin combine the flux of all zenith
angles. Table~\ref{tab:error} shows the flux, statistical and
systematic uncertainty for all zenith angle and energy bins. The
expected SSM flux of a particular energy bin is calculated from the
total $^8$B and {\it hep} flux of BP2000\cite{BP2000} and the neutrino
spectrum from Ortiz {\it et al.}\cite{newb8}.  This neutrino spectrum
is based on an improved measurement of the $\beta$-delayed $\alpha$
spectrum of the $^8$B decay with a small and well-controlled
systematic uncertainty.  Earlier reports\cite{SKspec} used the
neutrino spectrum by Bahcall {\it et al.}\cite{BP96}.

This zenith angle spectrum is analyzed in  a two-neutrino
oscillation scenario, which can be described with a mixing angle
$\theta$ and a mass difference $\Delta m^2$. We consider two cases:
(i)  $ \nu _e \leftrightarrow \nu _{\mu,\tau} $ and 
(ii) $ \nu _e \leftrightarrow \nu _{\mbox{\scriptsize sterile}}$.
For each set of neutrino oscillation parameters ($\sin^2 2\theta$
and $\Delta m^2$) the expected number of solar neutrinos and its
zenith angle spectrum are calculated. First, the probability
$P_1$ ($P_2$) of a solar neutrino to be in the mass eigenstate $\nu_1$ ($\nu_2$)
on the surface of the sun is obtained from a numerical calculation
which propagates
a neutrino wave function from the production point in the core to the
surface. This calculation uses models for the distribution of the neutrino
production point in the sun~\cite{BP2000}, the electron density
in the sun~\cite{BP2000}, and the neutrino
spectrum~\cite{newb8}.
Above $\Delta m^2=1.8\cdot10^{-9}$ eV$^2$ the propagation of the
two mass eigenstates from the sun to the earth and inside the earth
can be assumed to be incoherent. The survival probability at the
detector is given by
\begin{equation}
  P(\nu_e \rightarrow \nu_e)_{SK} = P_1P_{1e} + P_2P_{2e},
  \label{eq:survsk}
\end{equation}
where $P_{1e}$ ($P_{2e}$) is the probability to be $\nu_e$ at the detector
if the neutrino arrives at the earth as $\nu_1$ ($\nu_2$)
taking into
account matter effects inside the earth~\cite{MSW}.
The electron density model for the earth (PREM~\cite{PREM})
assumes a charge-to-mass ratio (Z/A) of 0.468 for the core and 0.497
for the mantle~\cite{mantleratio}.
Below $\Delta m^2=1.8\cdot10^{-9}$ eV$^2$ matter effects inside the earth are
unimportant and the propagation from the sun to the earth is assumed
to be coherent. The survival probability is then
%
\begin{eqnarray}
  P(\nu_e \rightarrow \nu_e)_{SK} &=& P_1\cos^2\theta + P_2\sin^2\theta +\cr
& & 2\times\sqrt{P_1P_2}\cos\theta\sin\theta
  \cos(\frac{\Delta m^2L}{2E_{\nu_e}}),
  \label{eq:surv}
\end{eqnarray}
%
where $L$ is the distance from the sun to the earth
ranging from perihelion (winter) to aphelion (summer)
\footnote{In the case of vacuum oscillations
$P_1=\cos^2\theta, P_2=\sin^2\theta$, and this reduces to
$1-\sin^22\theta\sin^2\frac{\Delta m^2L}{4E}$}.

\begin{figure}[tb]
\centerline{\includegraphics[width=9cm]{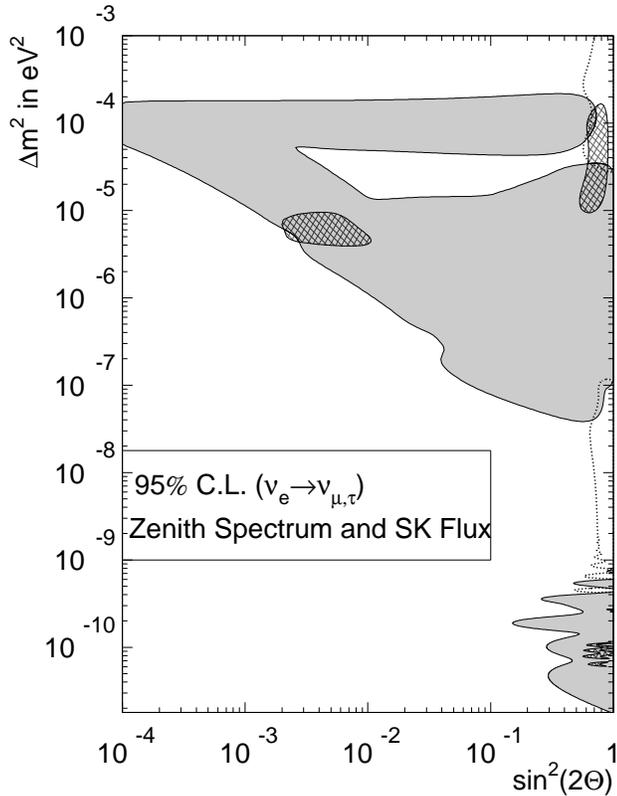}}
\caption{Exclusion area for two-flavor oscillation
$\nu_e\leftrightarrow\nu_\mu/\nu_\tau$
from zenith angle spectrum analysis at 95\%
confidence level. Overlaid are the allowed areas (95\% C.L.)
using the zenith angle spectrum and the SSM flux
prediction (dotted lines). The small overlap of allowed area
and excluded area is due to the fairly good agreement of the flux
for these parameters.
The hatched areas are allowed at 95\%
in a combined fit to the fluxes measured at GALLEX\protect\cite{Gallex},
SAGE\protect\cite{SAGE}, Homestake\protect\cite{Cl} and
Super-Kamiokande\protect\cite{SKflux2}.}
\label{activearea}
\end{figure}

\begin{figure}[tb]
\centerline{\includegraphics[width=9cm]{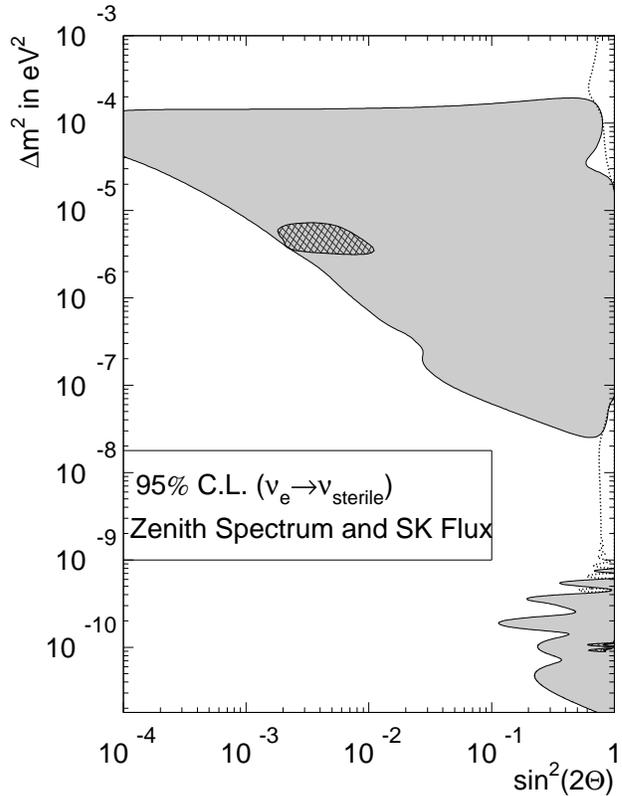}}
\caption{Exclusion area for two-flavor oscillation
$\nu_e\leftrightarrow\nu_{\mbox{\scriptsize sterile}}$
from zenith angle spectrum analysis at 95\%
confidence level. Overlaid are the allowed areas (95\% C.L.)
using the zenith angle spectrum and the SSM flux
prediction (dotted lines). The hatched areas are allowed at 95\%
in a combined fit to the fluxes measured at GALLEX\protect\cite{Gallex} and
SAGE\protect\cite{SAGE}, Homestake\protect\cite{Cl} and
Super-Kamiokande\protect\cite{SKflux2}.}
\label{sterilearea}
\end{figure}

Four experiments have quoted measurements of the solar neutrino flux:
Homestake~\cite{Cl} ($2.56\pm 0.16\pm 0.16$ SNU),
SAGE~\cite{SAGE} ($75.4^{+7.8}_{-7.4}$ SNU), GALLEX/GNO~\cite{Gallex}
($74.1^{+6.7}_{-6.8}$ SNU), and
SK~\cite{SKflux2}.
Taking an average of SAGE and GALLEX, 
a combined analysis results in several allowed regions.
The combined analysis is performed by the method given
in~\cite{methglob} considering
updated theoretical correlated and uncorrelated uncertainties
($^{37}$Cl cross section~\cite{BP96},
$^{71}$Ga cross section~\cite{crossga},
neutrino-electron scattering cross section~\cite{crossscat},
and diffusion~\cite{diffusion}).
The hatched areas of Fig.~\ref{activearea} show the allowed regions at
95\% C.L.($\chi^2 < \chi^2_{min}+5.99$)
assuming $\nu_e \rightarrow \nu_{\mu,\tau}$ oscillation.
There are four allowed regions
called ``small mixing angle solution''
(SMA, $\Delta m^2\approx10^{-5}$ eV$^2$,
$\sin^22\theta\approx10^{-2}\dots10^{-3}$),
``large mixing angle solution''
(LMA, $\Delta m^2\approx10^{-4}\dots10^{-5}$ eV$^2$,
$\sin^22\theta>0.5$),
``low solution''
(LOW, $\Delta m^2\approx10^{-7}$ eV$^2$,
$\sin^22\theta\approx0.9$)\footnote{This solution
only appears at 99\% C.L.; however it is usually discussed as
a possible solution.},
and ``just-so solution'' ($\Delta m^2<10^{-9}$ eV$^2$).
The SMA and just-so solutions predict spectral distortion, while the
LMA and LOW solutions predict a zenith angle variation and can therefore
be constrained by the zenith angle spectrum data of SK.
The hatched areas of Fig.~\ref{sterilearea} are the combined
analysis allowed regions at 95\% C.L. assuming
$\nu_e \leftrightarrow \nu_{\mbox{\scriptsize sterile}}$ oscillation.
In this case, LMA and LOW solutions do not occur, since there is
not enough neutral current contribution to neutrino-electron scattering
to accommodate the difference between the SK and the Homestake
flux results.

Using the zenith angle spectrum (see Fig.~\ref{daynightspec})
the probability of two neutrino oscillation scenarios was tested with a
$\chi^2$ method. For each energy bin $i$, we form a zenith angle 
flux difference vector
$\overrightarrow{\Delta_i}$. Its seven zenith components $\Delta_{i,z}$ are
\[
\Delta_{i,z}=
\frac{\phi^{\mbox{\tiny meas}}_{i,z}}{\phi^{\mbox{\tiny SSM}}_i}-
\alpha\times f\left(E_i,\delta_{\mbox{\tiny corr}}\right)\times
\frac{\phi^{\mbox{\tiny osc}}_{i,z}}{\phi^{\mbox{\tiny SSM}
}_i}
\]
where $z$ is the zenith angle bin,
$\phi_{i,z}^{\mbox{\tiny meas}}$ is the observed flux of each
energy and zenith angle bin, and  $\phi^{\mbox{\tiny SSM}}_i$ and
$\phi^{\mbox{\tiny osc}}_{i,z}$ are the
expected event rates in that bin without and with neutrino
oscillation. The spectral distortion $f$ due to the correlated
systematic error (see table~\ref{tab:error})
of $\phi_{i,z}$ is scaled by the parameter
$\delta_{\mbox{\tiny corr}}$.
The definition of the $\chi^2$ is thus
\begin{equation}
  \chi^2 = \sum_{i=1}^{8}
\overrightarrow{\Delta_i}\cdot V_i^{-1}\cdot\overrightarrow{\Delta_i}
+\left(
\frac{\delta_{\mbox{\tiny corr}}}{\sigma_{\mbox{\tiny corr}}}
\right)^2
\end{equation}
Each energy bin $i$ has also a separate $7\times7$
error matrix $V_i$ describing the energy-uncorrelated uncertainty.
$V_i$ is the sum of the statistical error matrix and the
energy-uncorrelated systematic error matrix (see table~\ref{tab:error}),
the latter of which is
constructed
assuming full correlation in zenith angle.
The flux normalization factor
$\alpha$ is unconstrained to make the $\chi^2$ independent
of the total solar neutrino flux. The correlation parameter
$\delta_{\mbox{\tiny corr}}$ is constrained
within $\sigma_{\mbox{\tiny corr}}$.
The size and shape of the correlated error are calculated as in \cite{SKspec}.
The {\it hep} contribution to the neutrino flux is not constrained.

The $\chi^2$ values are calculated in the parameter space,
($10^{-4}\leq\sin^2 2\theta\leq1$,
$10^{-11}$ eV$^2\le\Delta m^2\leq10^{-3}$ eV$^2$).
In the case of active neutrinos, the minimum $\chi^2$ value is
36.1 with 40 degrees
of freedom at ($\sin^2 2\theta = 1$,
$\Delta m^2 = 6.53\cdot10^{-10}$ eV$^2$).
The best-fit flux normalization is $\alpha=0.788$,
the correlation parameter
is $\delta_{\mbox{\tiny corr}}=-0.06\sigma_{\mbox{\tiny corr}}$ and
the {\it hep} flux is $0$.
The shaded areas in
Fig.~\ref{activearea} are
excluded at 95\% C.L. from this flux independent analysis.
Most of the SMA and just-so solutions are disfavored with this C.L..
In the case of sterile neutrinos, the minimum $\chi^2$ value is 35.7
at ($\sin^2 2\theta=1$,
$\Delta m^2=6.57\cdot10^{-10}$ eV$^2$). 
All possible solutions are disfavored at 95\% C.L. in this case.
The best-fit flux normalization is $\alpha=0.917$, the correlation parameter
is $\delta_{\mbox{\tiny corr}}=0.06\sigma_{\mbox{\tiny corr}}$ and
the {\it hep} flux is $0$.
The shaded areas in
Fig.~\ref{sterilearea} show the excluded regions (95\% C.L.).

Using the theoretical uncertainty of the $^8$B flux
$\sigma_{\mbox{\tiny flux}}=^{+0.20}_{-0.16}$SSM, an analysis
combining flux and zenith angle spectrum
has also been performed.
%
%
In the active neutrino case, the minimum $\chi^2$ value is 37.8 with 41 d.o.f.
at the same position as the unconstrained case.
The flux normalization changes to $\alpha=0.789$ and
the correlation parameter to
$\delta_{\mbox{\tiny corr}}=-0.02\sigma_{\mbox{\tiny corr}}$
The minimum $\chi^2$ point is within the just-so solution,
but some LMA $\chi^2$ are similar to the minimum.
For example, $\chi^2 = 39.1$ at ($\sin^2 2\theta = 0.87, \Delta m^2 =
7\cdot10^{-5}$ eV$^2$) with a {\it hep} flux of 2.9$\times$BP2000.
The dotted lines in
Fig.~\ref{activearea} show the contours of the 95\% C.L.
allowed regions.
In the sterile neutrino case, the minimum $\chi^2$ value is 35.9
with 41 d.o.f at the same position as the unconstrained case.
Flux normalization, correlation
parameter and {\it hep} flux are unchanged.
The inside of dotted lines in
Fig.~\ref{sterilearea} is allowed at 95\% C.L.. Since the allowed area
from the combined flux analysis does not overlap these regions, oscillations
into only sterile neutrinos are disfavored at this confidence level.

\begin{figure}[tb]
\centerline{\includegraphics[width=8.8cm]{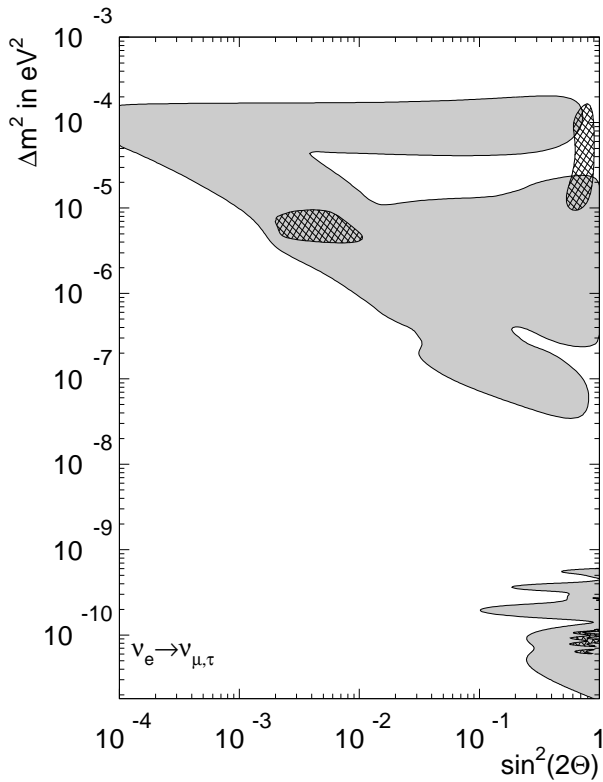}}
\caption{Exclusion area for two-flavor oscillation
$\nu_e\leftrightarrow\nu_\mu/\nu_\tau$ from a day/night spectrum analysis.
The hatched area is allowed at 95\%C.L. by
the combined flux analysis same as in Fig.~\ref{activearea}
}
\label{bahcallspec}
\end{figure}



Figures~\ref{activearea} and \ref{sterilearea} are based on the
$\chi^2$ analysis of the zenith angle spectrum.  We have also
performed an oscillation search using the ``day/night spectrum'',
which, in contrast to the zenith angle spectrum, divides the data into
two zenith angle bins (day and night bin).  Each of these bins is then
divided into 19 energy bins~\cite{SKflux2}.  The $\chi^{2}$ is defined
as follows: \footnotesize
\begin{eqnarray*}
\lefteqn{\chi^2 = \sum_{D,N} \sum_{i=1}^{19}} \\
 & & \left(
\left(
\frac{\phi^{\mbox{\tiny meas}}_{i,D/N}}{\phi^{\mbox{\tiny SSM}}_i}-
\alpha\times f\left(E_i,\delta_{\mbox{\tiny corr}}\right)\times
\frac{\phi^{\mbox{\tiny osc}}_{i,D/N}}{\phi^{\mbox{\tiny SSM}}_i}
\right)
/\sigma_i \right)^2 \\
 & &  +\left(
\frac{\delta_{\mbox{\tiny corr}}}{\sigma_{\mbox{\tiny corr}}}
\right)^2,
\nonumber
\end{eqnarray*}
\normalsize The notation is analogous to that used in the $\chi^{2}$
definition of the zenith angle spectrum analysis.  $\sigma_{i}$ is the
sum of statistical and uncorrelated errors added in quadrature.

 The minimum $\chi^2$ value is 28.2 with 34 degrees of freedom at $\sin^2
2\theta = 0.4$ and $\Delta m^2 = 1.38\cdot10^{-10}$ eV$^2$.  The
best-fit flux normalization is $\alpha = 0.488$ and the correlation
parameter is $\delta_{\mbox{\tiny corr}} = -0.2\sigma_{\mbox{\tiny
corr}}$.  Figure~\ref{bahcallspec} shows the 95\% excluded regions
using the shape of this day/night spectrum.  The excluded area is
similar to that obtained in the zenith angle spectrum, but more
restrictive in the SMA region.  The differences at the LMA and near
the LOW solution are due to the zenith angle variations within the
night bin.  The lower left corner of the SMA predicts a slight
depression of the core flux resulting in a day flux prediction that is
larger than the night flux.  SK measures a $1.3\sigma$ excess of the
night flux over the day flux\cite{SKflux2}, but the flux in the core
bin is below the day flux.  This leads to a slightly better fit of
these parameters to the zenith angle spectrum than to the day/night
spectrum.  The lower left corner of the SMA 95\% C.L. region is
excluded at 93\% C.L by the zenith angle spectrum and at 97\% C.L. by
the day/night spectrum analysis.  Other differences are due to the use
of different binnings.

In summary, Super-Kamiokande precisely measured the energy dependence
and zenith angle dependence of the solar $^8$B neutrino flux. The
data do not show a significant distortion of the spectrum or
zenith angle variation. This places strong constrainins on
neutrino oscillation solutions to the solar neutrino
problem independently of the flux expectation. If oscillations into active
neutrinos are assumed, just-so and the SMA solutions
are disfavored at 93\% (zenith angle spectrum) to 
97\% C.L.(day/night spectrum) 
and the LMA solutions are preferred.
In conjunction with the SK $^8$B flux measurement, two allowed areas
at large mixing remain.
All possible oscillation solutions into only sterile neutrinos are disfavored
at 95\% confidence level.

The authors acknowledge the cooperation of the Kamioka Mining and
Smelting Company.  The Super-Kamiokande detector has been built and
operated from funding by the Japanese Ministry of Education, Culture,
Sports, Science and Technology, the U.S. Department of Energy, and the
U.S. National Science Foundation.  This work was partially supported
by the Korean Research Foundation (BK21) and the Korea Ministry of
Science and Technology.

\end{document}